\documentclass[prl, 12pt]{revtex4-2}
\usepackage{amsmath}
\usepackage{amsfonts}
\usepackage{amssymb}
\usepackage{graphicx}

\usepackage{bm}
\usepackage{lmodern}
\usepackage{xspace}

\usepackage{color}

\newcommand{\mol}[1]{\ensuremath{_{\text{#1}}}}
\newcommand{\cmol}[2]{\ensuremath{_{\text{#1}}^{\text{#2}}}}
\newcommand{\mr}[1]{\ensuremath{\mathrm{#1}}}
\renewcommand{\v}[1]{\ensuremath{\bm{#1}}}
\newcommand{\zazb}{\ensuremath{Z_A/Z_B}\xspace}
\newcommand{\tss}[1]{\ensuremath{^{\text{#1}}}}

\newcommand{\pso}{\ensuremath{\left|\Psi_\mathrm{SO}\right \rangle}\xspace}
\newcommand{\pdo}{\ensuremath{\left|\Psi_\mathrm{DO}\right \rangle}\xspace}
\newcommand{\baco}{Ba\mol{2}CuO\mol{3}\xspace}
\newcommand{\bacoNNC}{Ba\cmol{2}{NNC}CuO\mol{3}\xspace}

\begin{document}

\title{Emergence of competing electronic states from non-integer nuclear charges}

\author{James W. Furness}\email{jfurness@tulane.edu}
\thanks{These authors contributed equally to this work}
\author{Ruiqi Zhang}
\thanks{These authors contributed equally to this work}
\author{Jamin Kidd}
\author{Jianwei Sun}
\email{jsun@tulane.edu}
\affiliation{Department of Physics and Engineering Physics, Tulane University, New Orleans, LA 70118, USA}

\date{\today}

\begin{abstract}

Understanding many-electron phenomena with competing near-degenerate electronic states is of fundamental importance to chemistry and condensed matter physics. One of the most significant challenges for exploring such many-electron phenomena is the necessity for large system sizes in order to realize competing states, far beyond those practical for first-principles methods. Here, we show how allowing non-integer nuclear charges expands the space of computationally tractable electron systems that host competing electronic states. The emergence of competing electronic states from non-integer nuclear charges is exemplified in the simple 2-electron H\mol{2} molecule and used to examine the microscopic structure of doped quasi-1D cuprate chains, showing how non-integer nuclear charges can open a window for first-principles calculations of difficult many-electron phenomena.

\end{abstract}


\maketitle

Complex materials are those exhibiting many-electron physics emerging from competition between intertwined charge, spin, orbital, and lattice degrees of freedom \cite{Fradkin2015}. While scientifically interesting in their own right, such materials present enormous technological potential as high-temperature superconductors and materials with charge-density wave orders. Understanding the competition between intertwined states remains a key step in controlling these desirable properties and their evolution under external controls like doping, temperature, pressure, and electromagnetic field.

Effective Hamiltonian models such as the Hubbard model \cite{Arovas2021} have been important tools in these studies, e.g., for cuprate superconductors \cite{Zheng2017, Huang2019}. These effective Hamiltonian models provide a generic description of competing electronic states that can be extended to better model specific classes of systems. For example, the 1D Hubbard model with the on-site Coulomb repulsion can be extended to include a near-neighbor attraction to better match angle-resolved photoemission data for quasi-1D cuprate chains \cite{Chen2021}. 

First-principles electronic structure theories give an alternative and complementary approach to effective Hamiltonian models by providing a material-specific model that addresses electrons directly with quantum mechanics. While in principle the first-principles approach can be exact for the electron degrees of freedom, the presence of many competing electronic states often makes their solution for complex materials intractable. For example, the parent compounds of cuprate high-temperature superconductors are anti-ferromagnetic (AFM) insulators which become metallic when doped with holes or electrons and then superconductive when further cooled \cite{Bednorz1986}. First principles models promise a window into how electronic states of cuprates evolve with doping and how competition between them emerges, though the size of interesting systems has so far made such models impractical.

Here, we show that allowing the nuclear charges to become non-integer drastically expands the space of systems that host competing electronic states, uncovering complex materials small enough to allow first-principles explorations. Using non-integer nuclear charges (NNC) we show competing strongly-correlated and charge-transfer states emerge from the simple 2-electron H\cmol{2}{NNC} molecule, and how the competition between these states can drive exotic phase transitions when H\cmol{2}{NNC} units are arranged into a stretched 1D chain. From this theoretical starting point we apply the NNC approach in a systematic first-principles investigation of the competing electronic states in quasi-1D cuprate chains under doping \cite{Chen2021}.


The H\mol{2} molecule is a paradigm system for strong correlation with a spin-singlet ground state that localizes the one electron onto each proton at dissociation. The symmetric nuclear potential from the equivalent hydrogen nuclei cannot exhibit charge transfer however, as both nuclei exert an equal pull on the two electrons. This left-right symmetry can be broken by replacing one proton with a helium nucleus  enabling charge transfer. This drives both electrons to localize on the heavier He nucleus at dissociation, and suppresses strong correlation. This complete replacement of strong correlation by charge transfer is unfortunate since it is the competition between them drives exotic properties in many materials \cite{Zaanen1985}. 

\begin{figure}
    \centering
    \includegraphics[width=\textwidth]{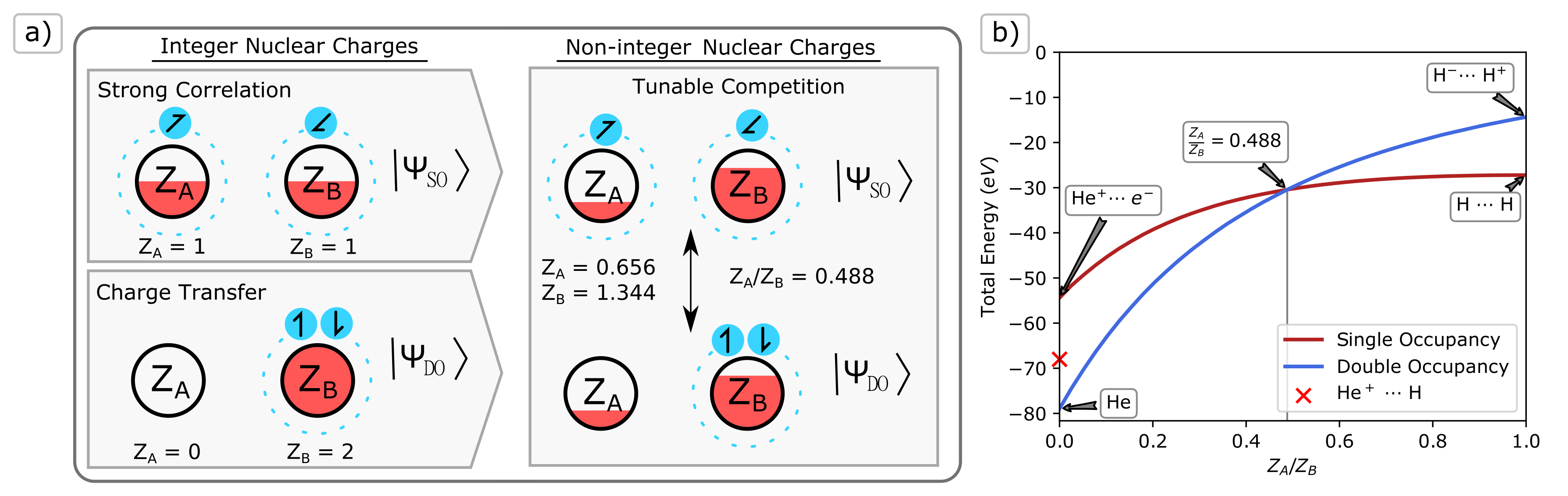}
    \caption{\textbf{Competition between charge transfer and strong correlation illustrated by H\cmol{2}{NNC} at infinite separation.} (a) A schematic plot of the single occupancy (SO) $\left|\Psi_\mathrm{SO}\right \rangle$, and double occupancy (DO) $\left|\Psi_\mathrm{DO}\right \rangle$ configurations in the spin-singlet state for integer and non-integer nuclear charges. The nuclear charges $Z_A$ and $Z_B$ are constrained such that $Z_A + Z_B = 2$ with non-integer charge represented by the red area within a circle of area 2. The blue discs represent electrons with spins labeled as arrows with tilted arrows denoting spin up-down degeneracy. (b) Total energies for SO and DO configurations as a function of the nuclear charge ratio, $Z_A/Z_B$, calculated with the coupled cluster method at the singles-doubles level (CCSD). Within annotations, ``$\cdots$'' denotes infinite separation. These CCSD calculations consider every possible excitation and are exact energies for the large d-aug-cc-pV5Z basis set \cite{Woon1994}. Turbomole version 7.4.1 was used for CCSD calculations with non-integer nuclear charges \cite{Hattig2010, Turbomole2020}.}
    \label{fig:frac_atoms}
\end{figure}

By allowing the hydrogen nuclei to have non-integer charges, the NNC H$_2$ molecule (denoted as H\cmol{2}{NNC}) extends H$_2$ to a new paradigm system exhibiting competition between strong correlation and charge transfer states. To make the system charge neutral we choose that the two nuclear charges, $Z_A$ and $Z_B$, sum to $+2$ and we require $Z_A \leq Z_B$ without loss of generality. Tuning $Z_A/Z_B$ away from 1 enables charge transfer without completely suppressing strong correlation. This H\cmol{2}{NNC} system is shown schematically in Figure \ref{fig:frac_atoms} a).

There are two possible electronic configurations for the H\cmol{2}{NNC} system at infinite nuclear separation: a single occupation (SO) solution, \pso, with one electron on each nucleus characterizing strong correlation, and a double occupation (DO) solution, \pdo, with both electrons on the more charged $Z_B$ nucleus characterizing charge transfer. Depending on the ratio $Z_A/Z_B$, the ground state is either the \pso or \pdo, or the configurations are degenerate. This is summarized in Figure \ref{fig:frac_atoms} a), while Figure \ref{fig:frac_atoms} b) shows how the energies of the two configurations change across the range of $0 \leq Z_A/Z_B \leq 1$, resulting in a discontinuous ground state. At $Z_A/Z_B = 1$, the ground state \pso is comprised of separated neutral hydrogen atoms, while \pdo, a H\tss{-} ion and a proton, is higher in energy by around 12 eV. At $Z_A/Z_B = 0$, $Z_B$ becomes a Helium nucleus and $Z_A$ disappears so the converse is true: \pdo, a lone neutral helium atom, is favored over \pso, a He\tss{+} ion and a free electron. This localization of both electrons onto the more charged nucleus is similar to the dissociation behavior of HHe\tss{+}. Introducing non-integer nuclear charges forms a continuum between these limits with the energy of each configuration varying smoothly as the strong correlation of \pso competes with the charge transfer of \pdo. The \pso and \pdo configurations become degenerate at $Z_A/Z_B \approx 0.488$, at which point the strong correlation and charge transfer competition is maximized. 

The appearance of degeneracy between \pso and \pdo has profound implications when the charge-neutral H\cmol{2}{NNC} unit is extended to a (H\cmol{2}{NNC})$_\infty$ chain with uniform inter-nuclear distances. At short inter-atomic distances the $\zazb = 1$ hydrogen chain is weakly correlated and metallic, while at larger inter-atomic distances it undergoes a phase transition to a strongly correlated insulating phase \cite{Hachmann2006, Stella2011, Motta2017, Motta2020}, a prototypical example of the Mott--Hubbard metal-insulator transition \cite{Mott1974}. Consider this (H\cmol{2}{NNC})$_\infty$ chain with large inter-atomic distances such that the electron density overlap between atomic sites is negligible. Under these conditions the $\zazb = 1$ chain is an insulator. Now, following the previous analysis we can tune the \zazb non-integer nuclear charge ratio to bring each $Z_A-Z_B$ pair close to the SO-DO degeneracy. Around this point a small perturbation, e.g. an electric field that enhances the potential at the more positive nucleus of the pair, can drive an electron from a less positive to a more positive nuclear site. This charge transfer capability under small perturbation emerging from the insulating hydrogen-like chain highlights the rich physics brought by the non-integer nuclear charge.

\begin{figure}
    \centering
    \includegraphics[width=\textwidth]{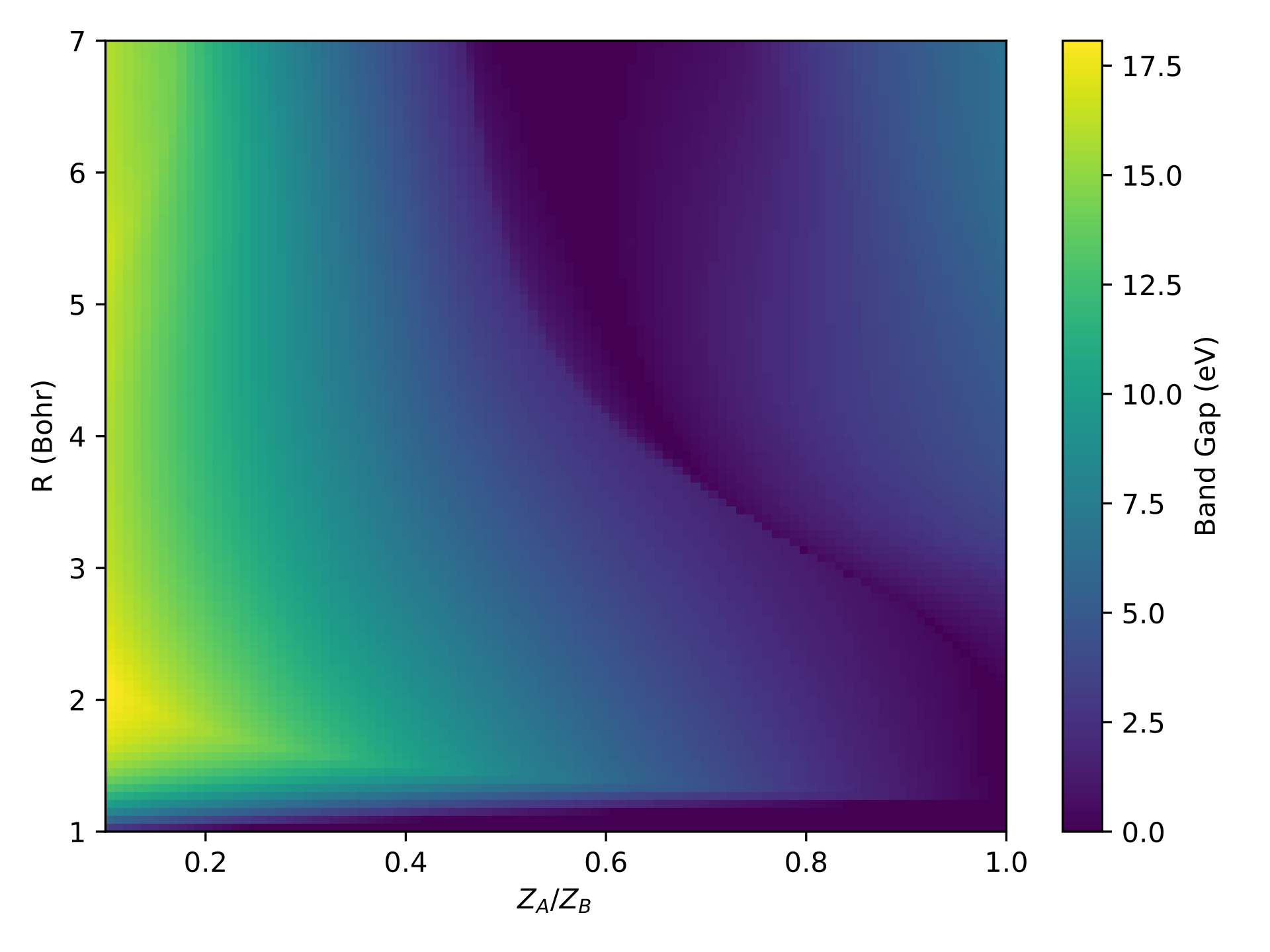}
    \caption{Band gap (eV) for the 1D (H\cmol{2}{NNC})$_\infty$ chain as a function of nuclear charge asymmetry $(Z_A/Z_B)$ and nuclear separation R, calculated by density functional theory using the SCAN exchange-correlation functional\cite{Sun2015}. The def2-TZVP Gaussian type basis set \cite{Weigend2005} was used for orbitals and density fitting in the continuous fast multipole method (CFMM). 10,000 $k$-space samples were taken along the periodic direction including the $\Gamma$ point and Gaussian smearing of orbital occupations was used with $\sigma = 0.001$. Each calculation was begun from a spin-symmetry broken set of guess orbitals and then optimized to self-consistency.}
    \label{fig:band_gap_landscape}
\end{figure}

Following this thought experiment we have carried out density functional theory (DFT) calculations of this (H\cmol{2}{NNC})$_\infty$ 1D chain, with all nuclei uniformly separated by distance $R$. The complexity of the infinite chain prevents exact diagonalization so we have used DFT with the strongly constrained and appropriately normed (SCAN) exchange-correlation density functional \cite{Sun2015} with broken spin-symmetry to determine the band gap for this system as a function of \zazb and $R$. DFT using the SCAN exchange-correlation functional is only an approximate solution of the electronic structure problem with a limited description of many electron effects, however it is a first-principles method that has been shown to make relatively accurate predictions for the band gaps of complex materials \cite{Furness2018, Lane2018, Zhang2019a, Zhang2020c, Lane2020a, Perdew2017, Perdew2021}. We defer further discussion of DFT, more accurate methods, and further technical details of these calculations to the Methods section.

The size of the (H\cmol{2}{NNC})$_\infty$ band gap is shown in Figure \ref{fig:band_gap_landscape} as a function of nuclear charge asymmetry, \zazb, and inter-nuclear separation, $R$. The right hand edge of this figure presents the well studied case of the 1D hydrogen chain, equivalent to (H\cmol{2}{NNC})$_\infty$ with $\zazb = 1$. In this limit we find a metallic character with vanishing band gap at small nuclear separations becoming insulating as nuclear separation increases, consistent with previous studies \cite{Hachmann2006, Stella2011,Motta2017, Motta2020}. This metallic character at small nuclear separation persists as nuclear charge symmetry is broken until both electrons localize onto the more charged $Z_B$ nucleus when $\zazb < 0.4$, suppressing conductivity for the range of $R$ considered. 

Following the top edge of $R = 7$ Bohr we see a distinct closing of the band gap around $\zazb \approx 0.55$, close to the degeneracy observed at $\zazb \approx 0.488$ for $R = \infty$ (Figure \ref{fig:frac_atoms} b). Consistent with our thought experiment, the near degeneracy of the strongly correlated single occupation \pso and charge transfer double occupation \pdo configurations removes the barrier for an electron to hop between sites shrinking the band gap towards zero. This transition is remarkable as it is totally driven by the competition between strong correlation and charge transfer, appearing at large nuclear separations where Mott insulator character would be expected for the normal 1D hydrogen chain. When $\zazb < 0.5$ however, both electrons in the unit cell localize onto the more charged $Z_B$ and a band insulator character is restored. 

The calculations of Figure \ref{fig:band_gap_landscape} qualitatively confirm our thought experiment, though the limitations of the DFT model (discussed in the Methods section) will likely introduce artifacts into the quantitative detail. The corresponding DFT band structures are presented in the supplemental material with this caveat in mind.

\begin{figure}
    \centering
    \includegraphics[width=0.85\textwidth]{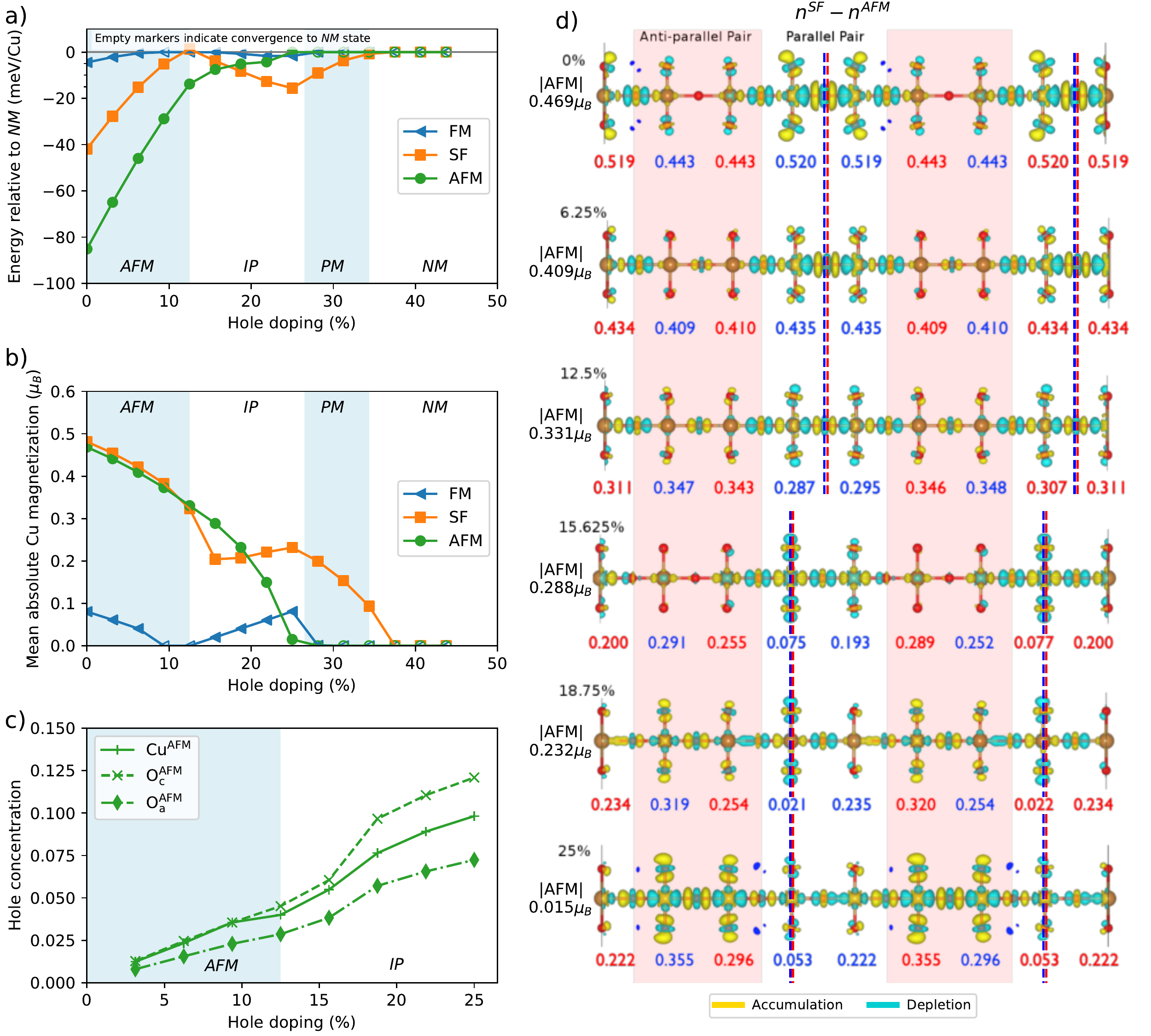}
    \caption{a) Total energy of the FM (blue), SF (orange), and AFM (green) spin configurations of \bacoNNC relative to the NM configuration, as a function of hole doping percentage. Calculations which converged to a non-magnetic state are indicated by empty markers. Vertical bands indicate anti-ferromagnetic (AFM), intertwined phase (IP), paramagnetic (PM), and non-magnetic (NM) regimes. The same regimes are indicated in the other sub-figures. Hole doping is modeled using the technique of non-integer nuclear charges discussed in the text. b) Mean absolute copper magnetization, and c) hole concentration in the AFM configuration evaluated by Bader hole population analysis~\cite{Tang_2009} for copper (solid), chain oxygen (dashed), and apical oxygen (dot-dashed), as a function of hole doping. d) Charge reorganization between SF and AFM configurations illustrated as the difference in electron charge-density. Yellow indicates accumulation and blue depletion. Copper atom magnetization in the SF configuration is given below each copper site, with red and blue text coloring indicating spin up and down character, and black ``$|$AFM$|$'' labels show absolute copper magnetic moment for the AFM configuration. Spin anti-parallel pairs are highlighted with red background shading, while spin-parallel pairs are unshaded. Vertical dashed lines indicate AFM domain boundaries in SF configurations. All calculations use the SCAN XC functional, and further calculation details are given in the Methdods section following the main text.}
    \label{fig:cu_1d}
\end{figure}


Now, we turn our attention to how non-integer nuclear charges can be connected to real materials through the lens of the prototypical quasi-1D cuprate chain: \baco~\cite{Chen2021}. This material and related hole-doped structures have been the focus of much recent interest, with an experimental realization recently reported by Chen \textit{et al} in Ref. \citenum{Chen2021}. Alongside spectroscopic experiments, Chen \textit{et al} find that a simple Hubbard model Hamiltonian requires addition of a sizable near-neighbor attraction to capture the spectroscopic features observed. Electron-electron Coulomb interactions are necessarily repulsive in nature so Ref.  \citenum{Wang2021} assigns this attractive term to coupling between electron and apical oxygen phonon modes in a Hubbard--extended-Holstein model. Aiming to complement these studies, we will show how a density functional theory model of \baco~extended to allow non-integer charged barium sites, termed \bacoNNC, can give an additional \emph{first principles} window into the electronic structure of this complex material, without empiricism.

The extension to \bacoNNC allows the direct modeling of hole-doped states in the 1D Cu-O chain using a super-cell representation. Holes are generated by removing electrons from the super-cell while the positive nuclear charges of all barium atoms in the super-cell are uniformly reduced by the fraction required to keep the overall charge neutrality, see supplemental materials for details. This super-cell system can support many different magnetic configurations, allowing insight into the rich physics of competitions between charge, orbital, and spin degrees of freedom. We note that the lattice degrees of freedom can be relaxed, but we chose not to in our calculations as our focus is on the emergence of competing electronic states. The magnetic configurations considered are antiferromagnetic (AFM), ferromagnetic (FM),  ``spin flip'' (SF), and nonmagnetic (NM), supported in a $2\times8\times1$ super-cell. These are shown in Figure S4 of the supplemental material. The SF configuration is of particular interest, containing mixed domains of AFM and FM orders. Unlike H\cmol{2}{NNC} and its infinite chain, no nuclear symmetry has been broken in this model, in particular for the 1D Cu-O chain. The low-energy physics pertaining to the 1D Cu-O chain is therefore expected to be well captured by this model. Similar to our previous analysis of the (H\cmol{2}{NNC})$_\infty$ chain, these calculations use the SCAN exchange-correlation functional. We note, in all systems considered here, the numbers of electrons in the super-cell are kept integer, as shown in Table S1 of Supplementary Materials, and thus high-level wavefunction and Green's function based methods can be applied in principle.

First, we focus on the pristine cuprate chain (0\% doping) where more experimental data are available. The left edge of Figure \ref{fig:cu_1d} a) shows that our first principles calculations predict the AFM configuration is more energetically stable than the other magnetic configurations considered, indicating an AFM magnetic coupling as observed experimentally \cite{Motoyama1996, Kim1996, Kim2006}. The band gap of the AFM configuration is predicted to be 1.12 eV, in agreement with experimental measurements of related Ca\mol{2}CuO\mol{3} (1.7 eV) and Sr\mol{2}CuO\mol{3} (1.5 eV) materials \cite{Maiti1998}.
Additionally, the Cu ions are predicted to have an average absolute magnetic moment of $0.469 \mu_\mr{B}$, close to the magnetic moments typically observed in cuprates \cite{Tranquada2007}. The $J$ exchange-coupling between the nearest Cu ions can be calculated from the total energies of the AFM and SF configurations, using the Heisenberg Hamiltonian,
\begin{equation}
	H = \frac{J}{2} \sum_{i} \v{S_i} \v{S_{i+1}}, \label{eq:heisenberg}
\end{equation}
where $\v{S_i}$ is the spin vector at site $i$, and $J > 0$ and $J < 0$ represent the AFM and FM spin exchange interactions respectively. Our calculations predict $J = 0.346$ eV, in good agreement with $J = 0.3$ eV obtained in Ref. \cite{Chen2021} by fitting to experimental data. This generally good agreement of first principles predictions with existing experimental data for the pristine AFM ground state indicates that the SCAN DFT model can be usefully accurate for this complex material.

Now considering the hole-doped material, Figure \ref{fig:cu_1d} a) shows the relative stability of each magnetic configuration with respect to the NM configuration as a function of hole doping percentage. The mean absolute magnetization of the copper atoms is also shown for each configuration as a function of hole doping in Figure \ref{fig:cu_1d} b). We can identify four phase domains from these figures. Up to $12.5\%$ hole doping, the AFM configuration is the most stable, defining an ``AFM regime". Throughout this regime increasing hole doping destabilizes all three configurations with respect to the NM configuration, accompanied by a decreasing copper magnetization. As the doping level increases from 12.5\% to 25\%, the three configurations become close in energy, to within 20 meV/Cu, indicating an ``intertwined phase'' (IP) regime. In particular, the energy differences between the AFM and SF configurations are smaller than 5 meV/Cu at 15.625\% and 18.75\% doping levels. Within this IP regime, some differences in magnetic response to increasing hole doping are seen: the magnetization of the copper sites in the AFM configuration continue to smoothly decline while for the SF and FM configurations copper magnetization increases. For the SF configuration this increase in magnetization is accompanied by a lowering of total energy resulting in the SF configuration becoming slightly favored over AFM, although all three phases remain close in energy. We note there is a clear drop in the average copper magnetization in the SF configuration when crossing from the AFM regime to the IP regime, indicating a first order phase transition. 

The stabilization of the SF, and to a lesser extent FM, configurations reaches a maximum around 25\% hole doping, after which a new paramagnetic regime (PM) is found where only the SF configuration retains the local magnetic moments. It is likely that more competing spin-disordered configurations exist around the IP and PM regimes, though for the present study we restrict ourselves to the relatively simple SF configuration supported by the $2\times8\times1$ super-cell system. In the PM regime, increasing hole doping lowers copper magnetization and destabilizes the SF configuration with respect to NM. By 35\% hole doping the copper magnetization is inhibited in all configurations and the system reverts to the NM configuration, defining the NM regime. 

Figure \ref{fig:cu_1d} c) shows how the dopant hole is distributed over the Cu, chain O, and apical O sites for the AFM configuration as a function of doping level. The hole density is reasonably equally distributed across the Cu and O sites, with each increasing at a similar rate with increasing doping level. For the copper sites this hole accumulation leads to the unpaired nominal $d$ electron number decreasing, resulting in the decrease in copper magnetization observed with hole doping. This reduction in magnetic moment has a destabilizing effect, raising the energy of the AFM configuration towards that of the NM configuration. At doping levels of $\leq12.5\%$ (AFM regime) the SF configuration shows similar behavior in magnetic moment and stability. In the $12.5\% - 25\%$ IP regime however, the behavior of the SF configuration is markedly different to the AFM configuration, with the average copper magnetization slightly increasing and energy decreasing as hole doping increases. 

To understand this different behavior Figure \ref{fig:cu_1d} d) plots the charge density difference between the SF and AFM configurations for selected doping levels, alongside the copper site magnetic moments for the SF configuration and mean absolute copper magnetization of the AFM configuration. The distribution of magnetic moments in the SF configuration reveals domains of AFM order (denoted AFM\mol{SF}) separated by boundaries, illustrated in Figure \ref{fig:cu_1d} d) with vertical dashed lines. When hole doping is $\leq 12.5\%$ this domain boundary lies on the chain oxygen between two Cu sites with parallel magnetic moments (bond centered). In the IP regime ($15\% - 25\%$) the domain boundary lies on one of these two Cu sites which had shared parallel magnetic moments at lower doping levels (site centered). This domain boundary switching from bond centered to site centered across the change from AFM to IP doping regimes aligns with the drop in mean absolute magnetization seen at the same point in Figure \ref{fig:cu_1d} b).

Studying the charge difference between SF and AFM configurations in Figure \ref{fig:cu_1d} d), the AFM doping regime ($\leq12.5\%$ hole doping) shows significant charge redistribution between the two configurations in the volume around the AFM\mol{SF} domain boundaries. This redistribution minimizes the effect of parallel neighboring magnetic moments on the AFM\mol{SF} domains, which is unfavorable at low doping levels, (see the relative energies of FM and AFM configurations in Figure \ref{fig:cu_1d} a)). 


A different pattern of electron redistribution is seen in the IP doping regime ($15\% - 25\%$), with significant reorganization within the AFM\mol{SF} domain and at the boundaries. This enhancement of magnetic moments in the AFM\mol{SF} domain and the diminishing of the magnetic moment at the boundary Cu site is a result of redistribution of the $d_{y^2-z^2}$ orbital among the two spin channels. Specifically, a Cu site with a spin-up magnetic moment within the AFM\mol{SF} domain changes spin-down into spin-up $d_{y^2-z^2}$ occupation, while this change is reversed for a Cu site at the AFM\mol{SF} boundaries. See supplemental materials Figure S5 for further details, in particular the one with the 18.75\% doping level. We note that the total net magnetic moment of the SF configuration is still zero. The above analysis shows that the stabilization of the SF configuration with respect to the AFM configuration in the IP regime results from the interplay between charge, orbital, and spin.       


Experimentally, Ref.~\cite{Chen2021} found that the holon folding bands observed at low doping levels disappear at $\sim 30\%$ hole doping level. Our prediction of a NM regime beyond $\sim 35\%$ doping level, where no local magnetic moment can be stabilized, offers a possible explanation for the disappearance of the holon folding band at $> 30\%$ hole doping, indicating the importance of spin degree of freedom. Our first principles calculations therefore suggest that the competing electronic states driven by the interplay between charge, orbital, and spin can be important in addition to the electron-phonon coupling~\cite{Wang2021} for deriving the microscopic understanding of the spectroscopic observations that are intensively pursued ~\cite{Chen2021, qu2022spin, wang2022spectral,tang2022traces}.

Although the precise pairing mechanisms for high-$T_{c}$ superconductors, including cuprates, still remain unclear, there is growing consensus that ``competing orders'' play a crucial role~\cite{Fradkin2015}. Our first-principles calculations show how such phenomena can emerge in a quasi-1D cuprate chain under hole doping.

In summary, from the above simple $H_2$ molecule and the prototypical \baco~1D cuprate chain under doping, the amenity of our non-integer nuclear charge model to first principles modeling shows how tuning competing electronic states in both finite and periodic systems can offer a handle to probe the electron behaviors of exotic materials.

\section{Methods}

\subsection{Non-Integer Nuclear Charges}

Non-Integer nuclear charges are implemented under the Born--Oppenheimer approximation by assigning desired $Z_i \in \mathbb{R}^+$ to each nucleus and evaluating the nuclear-electron attraction and nuclear repulsion integrals in the standard way. This modification is trivial for most existing electronic structure codes and is available in the standard \textsc{Turbomole} release used for this work \cite{Turbomole2020}. Within the NNC approach, the electrons are treated exactly in principle with quantum mechanics, while the nuclear charges come in as external parameters. The present application to the quasi-1D \baco~cuprate chain under doping is encouraging. In general an NNC model must be constructed carefully to connect to the underlying real material problem, in particular when nuclear degrees of freedom (including charge and mass) are important.

We note that other non-integer nuclear charge systems were investigated by Cohen and Mori-S\'anchez in the context of the DFT derivative discontinuity and delocalization error in Ref. \citenum{Cohen2014a} and by Zhang and Wasserman in the context of partition density functional theory in Ref. \citenum{Zhang2022}.

\subsection{Computational details \label{sec:Comp-details}}

All finite molecular calculations were carried out in the large d-aug-cc-pV5Z basis set, using the Turbomole program \cite{Turbomole2020}. Calculations of the (H\cmol{2}{NNC})$_\infty$ chain were performed using Turbomole with a periodic boundary condition along the bonding axis ($z$), and non-periodic dimensions perpendicular ($x, y$). This avoids the need for vacuum spacing between neighboring chain images that would be required for a calculation that is periodic in all three dimensions. The def2-TZVP Gaussian type basis set \cite{Weigend2005} was used for orbitals and density fitting in the continuous fast multipole method (CFMM). 10,000 $k$-space samples were taken along the periodic direction including the $\Gamma$ point and Gaussian smearing of orbital occupations was used with $\sigma = 0.001$. Each calculation was begun from a spin-symmetry broken set of guess orbitals and then optimized to a self-consistent symmetry broken ground state.

All calculations of \bacoNNC were performed by using the pseudopotential projector-augmented wave method \cite{Kresse1999} as implemented in the Vienna \emph{ab initio} simulation package (VASP) \cite{Kresse1993,Kresse1996}.
A high-energy cutoff of 520 eV was used to truncate the plane-wave basis set. The exchange-correlation effects were treated using the SCAN~\cite{Sun2015} density functional. A $7 \times 2 \times 5$ k-point mesh was used to
sample the Brillouin zone of these super-cells using the Monkhorst--Pack method. The electronic energy were converged to $10^{-6}$ eV. The structure parameters we used here are $a=3.827$\AA{}, $b=4.113$\AA{}, and $c=13.087$\AA{}, which were obtained by relaxing the AFM pristine phase with SCAN. The pseudo-potentials of Ba with non-integer nuclear charges were generated using code provided by VASP developers.

\subsection{Accuracy of methods used \label{sec:dft_accuarcy}}

First, the CCSD results of H\cmol{2}{NNC} in Figure \ref{fig:frac_atoms} are exact within the given basis set. The ability to calculate electronic structure exactly is invaluable for understanding the H\cmol{2}{NNC} system. Hartree--Fock (HF) theory is exact (for the chosen basis set) for single electron systems such as H and H\cmol{2}{+} but is insufficient for systems of multiple electrons. It is difficult in general to obtain exact solutions for multi-electron systems, normally requiring exponentially scaling methods such as full configuration-interaction (FCI). Fortunately, the two electron H\cmol{2}{NNC} system is small enough that exact diagonalization of the Fock space is feasible and a basis-set exact solution can be found. For technical convenience Figure \ref{fig:frac_atoms} uses the coupled-cluster method at the singles-doubles level (CCSD) which considers all possible excitations for a two electron system and is thus equivalent to exact diagonalization. We note however, that CCSD is not generally reliable for strongly correlated systems with more than two electrons.


Calculations of (H\cmol{2}{NNC})$_\infty$ and \bacoNNC used supercell representations with DFT employing the SCAN density functional.

DFT is exact in principle for the ground state and some excited-state \cite{perdew1985extrema} total energy and electron density in the Kohn-Sham (KS) \cite{Hohenberg1964, Kohn1965} or generalized KS (GKS) scheme \cite{Seidl1996}, if the exact XC functional were known. Within the GKS scheme, in which our calculations were conducted, the eigenvalue band gap of a density functional for a periodic system is equal to the functional's fundamental band gap, defined as the difference between ionization potential and electron affinity from total energy calculations \cite{Perdew2017}. If the exact XC functional were used then the ground states would be exact, giving an exact fundamental band gap. All the eigen-orbitals other than the frontier orbitals that define the eigen-energy band gap from a KS or GKS calculation are auxiliary however. Despite these formalities, in practice the accuracy of a DFT calculation is limited by the approximation to the exchange-correlation functional used.

Strongly correlated electron systems are generally challenging to DFT calculations using approximate exchange correlation functional although DFT being exact in principle. A common practice for accurately calculating total energies of such systems with approximate exchange correlation functionals is to allow symmetry breaking \cite{Perdew2021}. The appearance of broken symmetries of the electron density or spin density in a DFT calculation however can reveal strong correlations among the electrons that are present in a symmetry-conserved solution from high-level quantum mechanics methods \cite{Perdew2021}. For example, certain strong correlations present as fluctuations in the exact symmetry-conserved solution are frozen in symmetry-broken electron densities or spin densities of approximate DFT. This can be seen from the practical DFT solution of $H_2$ at a large nuclear separation, which breaks the spin symmetry for an accurate total energy by putting one spin-up electron on one proton and one spin-down on the other while the exact solution should be spin-singlet. In the infinite 1D chains studied here, the spin SU(2) symmetry should be conserved in the exact solutions if the spin-orbit coupling is not considered. In our DFT calculations with the SCAN density functional, we break the spin symmetry for better total energy descriptions by allowing local magnetic moments to be fixed in direction.

For the study of (H\cmol{2}{NNC})$_\infty$, SCAN can experience significant self-interaction error \cite{Fermi1934, Perdew1981} that causes the electron density to be overly diffuse, especially when the nuclear separation R is large. We expect this to manifest in (H\cmol{2}{NNC})$_\infty$ as a broadening of the vanishing band gap region of Figure \ref{fig:band_gap_landscape} by over-stabilizing the delocalization of the electrons across both centers. Despite this deficiency the SCAN calculations appear qualitatively correct, agreeing with our thought experiment about (H\cmol{2}{NNC})$_\infty$ at large separation. We have confirmed that SCAN shows qualitative agreement with exact FCI predictions of the fundamental gap of the H\cmol{2}{NNC} molecule, see supplemental materials Figure S1.

The (H\cmol{2}{NNC})$_\infty$ model in our DFT calculations has an equal nuclear separation and is represented by a periodic unit cell containing a single H\cmol{2}{NNC} unit. This limits exploration of some more complicated and exotic physics, including the short-range magnetic ordering, Peierls distortion \cite{Peierls2020}, etc.

Taking the 1D hydrogen chain, i.e., (H\cmol{2}{NNC})$_\infty$ with $\zazb = 1$, as an example, at the small nuclear separation the metallic phase can have short-range magnetic ordering \cite{Motta2020} which is inaccessible with our unit cell. We note however that band structures predicted by our SCAN calculations reveal the metal-insulator transition for the short-separation 1D hydrogen chain with the expected self-doping mechanism \cite{Motta2020}, see Supplemental Figure S2. At large nuclear separation a quasi-long-range anti-ferromagnetic (AFM) ordering has been predicted by density-matrix renormalization group (DMRG) for the equally spaced 1D hydrogen chain \cite{Motta2020}. Consistently, in our calculations, SCAN predicts the AFM configuration has the lowest energy. For (H\cmol{2}{NNC})$_\infty$ with $\zazb < 1$ at large nuclear separation, SCAN also finds the AFM configuration to be the lowest energy state associated with the band gap transition as $\zazb$ decreases from 1. Certainly, the AFM configuration being the lowest energy state here is the result of unit cell choice and the spin-symmetry breaking treatment of DFT calculations. We expect, however, that the total energy predicted by SCAN from this AFM configuration is accurate and close to the true ground state energy \cite{Perdew2021, Zhang2020c}. As these total energies are accurate, the generalized-Kohn--Sham band gap predictions of Figure \ref{fig:band_gap_landscape} are expected to be qualitatively correct for equally spaced (H\cmol{2}{NNC})$_\infty$ chains.  

While dimerization Peierls distortion \cite{Peierls2020} are important for hydrogen chain systems, we do not consider such effects here and constrain ourselves to chains with uniform inter-nuclear distances for the simplicity of being an illustration to our thought experiment.

Owing to the approximate nature of the DFT calculations and the limitations of small unit cell representation, the results of the (H\cmol{2}{NNC})$_\infty$ model should be confirmed with the high accuracy methods as in Refs. \citenum{Motta2017} and \citenum{Motta2020}, and with the periodic unit cell expanded to include more H\cmol{2}{NNC} units for more exotic phases. This is beyond the scope of the present investigation however. 

For the study of \bacoNNC, we expect the results predicted by SCAN are reasonably accurate. First, in comparison with conventional density functionals, SCAN has shown useful accuracy for many correlated materials \cite{Sun2016, Kitchaev2016, Furness2018, Lane2018, Zhang2019a, Zhang2020c}. In particular, SCAN has been demonstrated to provide accurate geometric, energetic, and magnetic properties for pristine \cite{Furness2018, Lane2018} and doped \cite{Zhang2019a} cuprates, although we note that SCAN is still an semilocal density functional approximation \cite{fu2018applicability, tran2020shortcomings,mejia2019analysis}. This improvement has in part been attributed to the reduction of self-interaction errors in SCAN \cite{Zhang2020c}. Second, SCAN predicts accurate magnetic and electronic properties for the pristine \baco~discussed in the main text. We expect more competing electronic states and physics will be discovered for \bacoNNC if larger supercells are used or many-electron techniques are applied on top of the electronic structures obtained with SCAN \cite{zhang2022fingerprints}.

\section{Acknowledgments}

J.W.F., R.Z., and J.S. acknowledge the support of the U.S. DOE, Office of Science, Basic Energy Sciences Grant No. DE-SC0019350. We thank John Perdew, Shaokai Jian, and Lin Hou for their comments.

\section{Author Contributions}

J.W.F. and J.S. conceived the non-integer nuclear charge approach. J.W.F., R.Z., and J.S. led the investigations, designed the computational approaches, and analyzed results. J.W.F. and R.Z. performed calculations. J.S. provided computational resources. J.W.F., R.Z., and J.S. wrote the manuscript. All authors contributed to editing the manuscript.

\section{Data Availability Statement}

Source data and input files are provided with this paper.

\section{Code Availability Statement}

Calculations of H\cmol{2}{NNC} and (H\cmol{2}{NNC})$_\infty$ use \textsc{Turbomole} version 7.4.1 \cite{Turbomole2020}.
Calculations of \bacoNNC use \textsc{VASP} version 6.2.1 \cite{Kresse1993, Kresse1996, Kresse1999}.

\section{Additional Information}
Correspondence and requests for materials should be addressed to \texttt{jfurness@tulane.edu} and \texttt{jsun@tulane.edu}. 

Reprints and permissions information is available at www.nature.com/reprints

\bibliography{2_electron_paradigms}

\end{document}